\documentclass[aps,twocolumn,groupedaddress]{revtex4-1}

\usepackage{graphicx}
\usepackage{dcolumn}
\usepackage{bm}
\usepackage{amsfonts}
\usepackage{amsmath}
\usepackage{amssymb}
\usepackage{xcolor}
\setcitestyle{super}
\usepackage{enumerate}

\def\hiroki#1{{\color{blue}#1}}

\begin{document}

\title{Disassembling of TEMPO-oxidized cellulose fibers: intersheet and 
interchain interactions in the isolation of nanofibers and unitary chains}



\author{Gustavo H. Silvestre$^1$, Lidiane O. Pinto$^2$, Juliana S. 
Bernardes$^{2,3}$, Roberto H. Miwa$^1$, and  Adalberto Fazzio$^{2,3}$}

\affiliation{$^1$Instituto de F\'isica, Universidade Federal de Uberl\^andia,
        C.P. 593, 38400-902, Uberl\^andia, MG, Brazil}
   
\affiliation{$^2$Brazilian Nanotechnology National Laboratory (LNNano), 
Brazilian Center for Research in Energy and Materials (CNPEM), Campinas, SP, 
13083-970, Brazil}

\affiliation{$^3$Center for Natural and Human Sciences, Federal University of 
ABC, Santo Andr\'e, S\~ao Paulo 09210-580, Brazil}

\date{\today}

\begin{abstract}

\textcolor{black}{Cellulose disassembly  is  an important issue  in designing 
nanostructures  using cellulose-based  materials. } In this work, we present a  
joint of experimental and theoretical study addressing the disassembly of  
cellulose nanofibrils. Through 2,2,6,6-tetramethylpiperidine-1-oxyl (TEMPO) 
mediated oxidation processes, combined with atomic force microscopy results, we 
find the formation of   nanofibers with diameters corresponding to a single 
cellulose  polymer chain. \textcolor{black}{The formation of these 
polymer chains is ruled by repulsive electrostatic interactions between the 
oxidized chains.} \textcolor{black}{Further first-principles 
calculations have been done in order to provide an atomistic understanding the 
cellulose disassembling processes,  focusing on the balance of  the interchain 
and intersheet interactions upon oxidation. Firstly we analyse these interaction 
in pristine systems, where we found  the intersheet interaction stronger 
than the  interchain  one.} In the oxidized systems,  we have considered the 
formation of (charged) carboxylate groups along the inner sites of elementary 
fibrils. We show  a net charge concentration  on the carboxylate groups, 
supporting the emergence of  repulsive electrostatic interactions between the 
cellulose \textcolor{black}{nanofibers}. Indeed, our total energy results show 
that the weakening of the binding strength between the fibrils is proportional 
\textcolor{black}{to the concentration and the}
net charge density of the carboxylate group. Moreover, by  comparing interchain 
and intersheet binding energies, we found that  most of the disassembly 
processes should take place by breaking the interchain O--H$\cdots$O hydrogen 
bond interactions, \textcolor{black}{and thus supporting the experimental 
observation of single and double cellulose polymeric chains.}

\end{abstract}

\maketitle

\section{Introduction}

Cellulose, the most abundant plant material resource, is a homopolymer with 
linear chains of glucopyranose rings linked through $\beta$-(1,4)-glycosidic 
bonds, which interact with other macromolecules as lignin and hemicellulose, 
within the plant cell 
walls\,\cite{taylor2008cellulose,mellerowicz2012tensional}. The search for more 
eco-friendly and energy-efficient technologies has accentuated the interest of 
using biomass to develop fuels, chemicals, and 
materials\,\cite{himmel2007biomass,jorgensen2007enzymatic}. Lately, nanoscale 
particles cellulose nanofibers (CNF) and cellulose nanocrystals (CNC) extracted 
from cellulose fibers by refinement have attracted attention as a 
next-generation material due to their outstanding mechanical 
properties\,\cite{klemm2011nanocelluloses,habibi2010cellulose}.

The resistance of plant biomass to breakdown and to fractionate into its 
molecular constituents is denominated 
recalcitrance\,\cite{himmel2007biomass,zhao2012biomass}. The biopolymers 
assemble in the cell wall, forming a robust microstructure hardy to physical, 
chemical, and enzymatic processes. Partially crystalline cellulose fibers are 
significant contributors to recalcitrance\,\cite{zoghlami2019lignocellulosic}. 
Thus, an understanding, at the electronic and atomic level, of the noncovalent 
interactions that bind the cellulosic chains is vital in designing effective 
refinement processes. 

\begin{figure}
    \includegraphics[width=7.5cm]{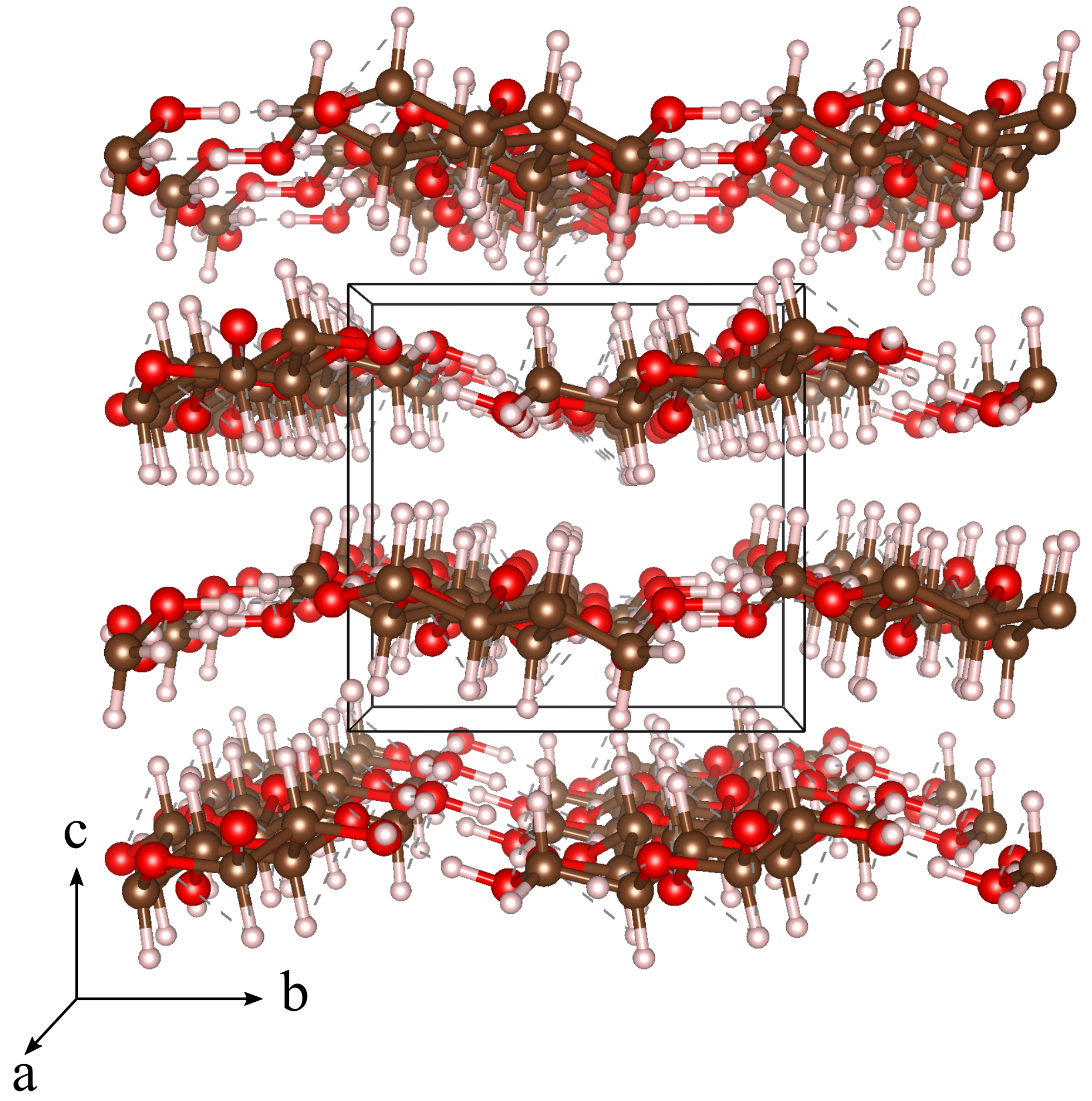}
    \caption{\label{fig:models} Structural model of monoclinic 
cellulose I$_\beta$. Solid lines show a perspective view of the periodic unit cell.}
  \end{figure}

The combination of synchrotron X-ray and neutron 
diffraction of cellulose I$\beta$, the mostly found allomorph of cellulose in 
higher plants, allowed to determine the most accurate position of all atoms in 
the unit cell, including the hydrogens\,\cite{nishiyamaJACS2002}. The 
crystalline structure consists of two parallel chains presenting slightly 
different conformations and organized into packed sheets 
(Fig.\,\ref{fig:models}). Both H-bondings within a single layer of cellulose 
(interchain interactions) and stacking interactions (intersheet interactions), 
mainly originated from vdW forces, contribute to stabilizing the cellulose 
crystal structure. However, there is no consensus in the  literature regarding  
which interaction is the key player in cellulose 
recalcitrance\,\cite{qianMacromolec2005,grossJPhysChemB2010,liJPhysChemC2011, 
devarajanJPhysChemB2013}.

Due to the compact and rigid structure of cellulose, size-reduction is very 
energy-intensive and demands harsh conditions. To extract nanofibers (CNFs), 
high-energy mechanical disintegration methods as refining, grinding, and 
homogenization are often used \cite{klemm2011nanocelluloses}. Chemical 
\cite{saito2007cellulose} and enzymatic treatments \cite{paakko2007enzymatic}, 
before mechanical action, are promising strategies to reduce the recalcitrance 
of cellulose. An established type of chemical pre-treatment consists of adding 
carboxylate groups (COO$^-$) on the surface of cellulose through TEMPO-mediated 
oxidation \cite{saito2007cellulose}. This surface modification is particularly 
interesting since it displays position-selective catalytic oxidation under 
moderate aqueous conditions. Electrostatic repulsion and osmotic effects acting 
between anionically-charged surfaces lead to the formation of individualized 
nanofibers after the mechanical process. In a recent work 
\cite{pintoCarbPol2019}, we observed that accomplishing TEMPO-mediated oxidation 
under a high concentration of NaClO, cellulose nanofibers from sugarcane bagasse 
were obtained without a mechanical defibrillation step. Besides, a significant 
number of nanofibers with diameters smaller than that corresponding to an 
elementary fibril ($<$ 3.5 nm) were imaged by atomic force microscopy (AFM), 
suggesting that besides promoting the disassembling of the fiber bundles into 
individual elementary nanofibers, the chemical oxidation also affects the 
intersheet and interchain interactions. 

In this work,  we performed a joint of experimental and theoretical 
investigation  of  cellulose disassembly upon oxidation, and the rule  played 
by the carboxylate groups  on the intersheet (IS) and interchain 
(IC) interactions along the inner sites of the CNFs. Firstly, through
TEMPO-mediated oxidation and based on  AFM measurements, we show the formation 
of  nanofibers with widths corresponding to the ones of  single and double 
cellulose polymer chains. In the sequence, based on first-principles 
calculations,  we present a total energy picture  of the  intersheet (IS) and 
interchain (IC) interactions along the (i) pristine and (ii) oxidized cellulose 
nanofibers. In (i), we have compared the IS and IC binding energies by using 
different approaches to describe the long-range van der Waals (vdW) 
interactions, and in (ii) we have examined the strength of the IS and IC 
interactions as a function of the degree of oxidation of the carboxylate groups.

\section{Materials and Methods}

\subsection{Experimental section}

\subsubsection{Cellulose Nanofibers Preparation}

CNFs were isolated by a procedure described elsewhere \cite{pintoCarbPol2019}. 
Briefly, the cellulose pulp extracted form sugarcane bagasse was 
surface-carboxylated by TEMPO oxidation reaction using \textcolor{black}{three 
volumes of a 12\% (w/v) NaClO solution (16.0, 78.0 or 156.0 mL per gram of 
cellulose). The average charge density was determined by three conductometric 
titration measurements, as previously described \cite{lin2012tempo}. Oxidized 
samples from sugarcane bagasse (SC) were identified as SC-5, SC-25 and SC-50, 
according to the concentration of NaClO used in the reaction (5, 25 or 50 mmol/g 
substrate).}

\subsubsection{Morphology characterization}

The morphology of oxidized CNFs was observed by atomic force microscopy (Park 
NX10) under ambient conditions and using the tapping mode. The cantilever from 
Nanoworld had a spring constant of 42 Nm$^{-1}$ and nominal resonance frequency 
of 75 kHz. Prior AFM analyses, a droplet of aqueous diluted CNF suspension 
(10\,$\mu$L, 5 mg/L) was deposited onto cleaved mica substrate (TED PELLA) and 
dried by natural evaporation at room temperature. \textcolor{black}{The length 
and height of CNFs were measured via Gwyddion 2.54 software by counting about 
300 independent nanofibers.}

\subsection{Computational details}

The calculations were performed by using the Density Functional Theory (DFT), as 
implemented in the computational codes  Quantum ESPRESSO (QE) software 
\cite{espresso}, and Vienna {\itshape ab initio} software package (VASP) 
\cite{vasp1, vasp2}. The exchange-correlation term was described within the 
generalized gradient approximation as proposed by Perdew, Burke and Ernzerhof 
(GGA-PBE) \cite{PBE}. The periodic boundary conditions were satisfied using the 
super-cell approach with a vacuum region in the direction perpendicular to the
cellulose sheet of at least 12 \AA\, to avoid image interactions. The 
Kohn-Sham (KS) orbitals were expanded in a plane wave basis set with an energy 
cutoff of 48\,Ry. We have verified the convergence of our results by 
increasing the energy cutoff up to 60\,Ry.  The electron-ion 
interactions were solved using the Projector Augmented Wave (PAW) method 
\cite{paw}, and the 2D Brillouin Zone (BZ) is sampled according to the 
Monkhorst–Pack method\,\cite{mp}, using a gamma-centered 3$\times$3$\times$1 
mesh for the cellulosic sheets and chains, and 3$\times$3$\times$3 for the 
cellulose crystalline phase. To determine the equilibrium configurations, the 
atomic positions and the lattice vectors were fully relaxed, considering a 
convergence criteria of  25\,meV\AA$^{-1}$ for the atomic forces on each atom, 
and  pressure smaller than  0.5\,Kbar.

In order to perform a thorough study of the role played by the  van der Waals 
(vdW) forces in the structural stability of  the cellulose nanofibers, we have 
taken into account different approaches for the vdW interactions, 
{\it viz.}: (i)  vdW density functional 
(vdW-DF)\,\cite{thonhauserPRL2015,thonhauserPRB2007,berlandRepProgPhys2015, 
langrethJPhysC2009} implemented in the QE and VASP codes, (ii) parameterized 
vdW-D2\,\cite{grimmeJCompChem2006} implemented in the QE code,  (iii)  
vdW-DF2\,\cite{hamada2014van}, and (iv) 
vdW-optB86b\,\cite{klimesJPhysC2010,klimevsPRB2011} both implemented in the VASP 
code.

\section{Results and Discussions}

\subsection{Experimental results}

\subsubsection{Overview of the cellulose nanofibers}

\begin{figure}
   \includegraphics[width=7.3cm]{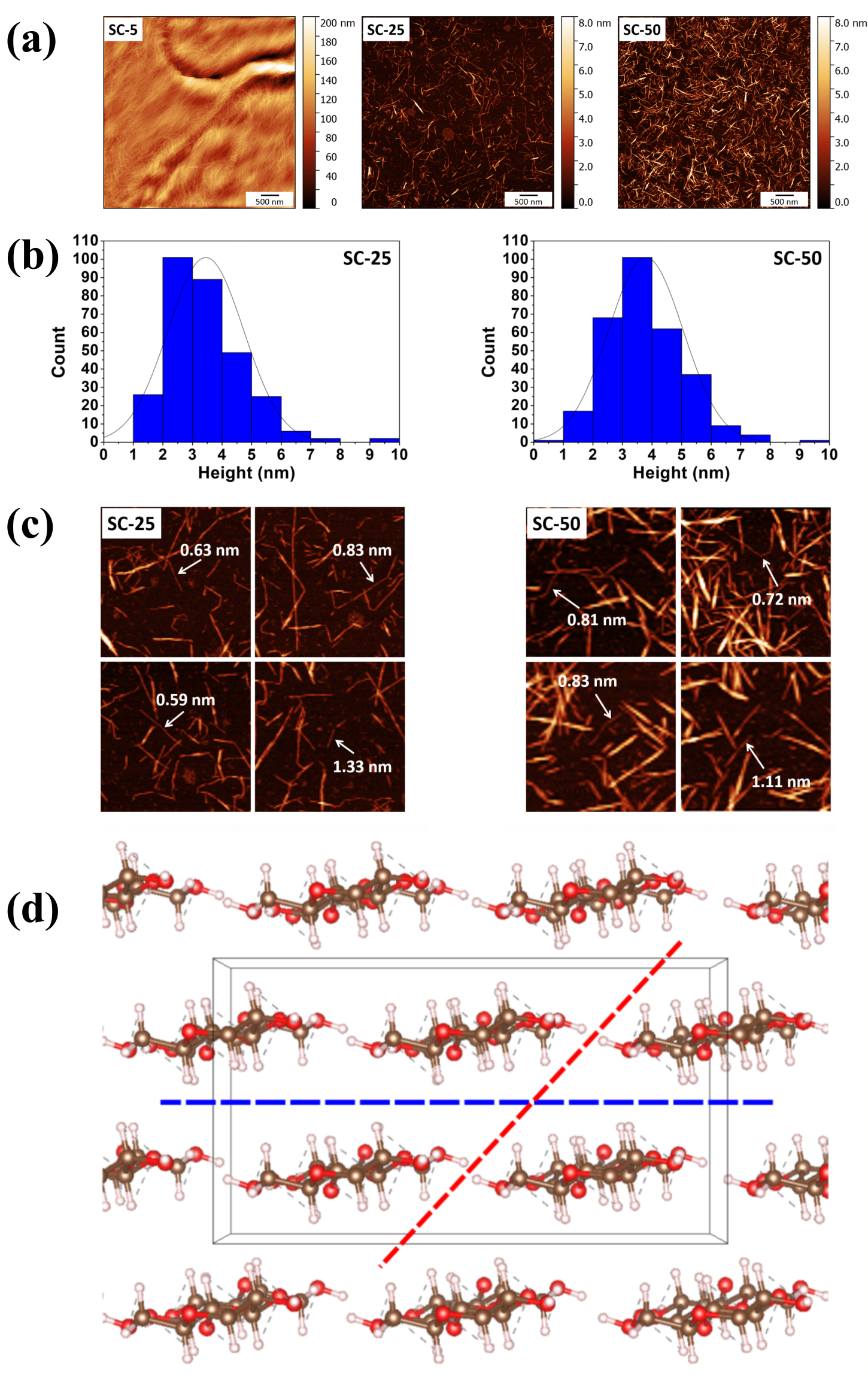}
    \caption{\label{fig:afm-results} Experimental results: AFM images (a) and 
(c), and elementary fibril diameter distribution (b). Representation of a inner 
sites of cellulose crystal (d). The red (black) dashed lines indicates a 
interchain (intersheet) disruption.}
  \end{figure}

The oxidized CNFs were isolated from sugarcane bagasse pulp through 
TEMPO-mediated oxidation using high oxidant content, 25 and 50\,mmol/g.  This 
reaction converts C6 primary hydroxyls groups from cellulose to carboxylates 
(COO$^{-}$) Na$^{+}$, yielding gravimetrically normalized values of 1.10 and 
1.40\,mmol of COO$^{-}$ per gram of cellulose, respectively, which correspond to 
ca 25\% of \textcolor{black}{oxidation}. Electrostatic repulsion between 
\textcolor{black}{highly}  charged cellulose microfibrils 
(\textcolor{black}{SC-25 and SC-50}$\zeta$-potentials ca $-65$\,mV in water) 
together with osmotic effects promoted the disassembling of completely 
individualized CNF dispersed in water without the need for high-energy 
mechanical treatments, as can be visualized in Fig.\,\ref{fig:afm-results}(a). 
\textcolor{black}{On the other hand, cellulose fibers with low carboxylate 
content (SC-5, 0.4 mmol per gram of cellulose) presented aggregated fibril 
bundles, as already observed for different types of biomass.}
The \textcolor{black}{SC-25 and SC-50} nanofibers present average lengths in 
the 
range of 243-370\,nm and an average width of 4\,nm, which may correspond to 
elementary fibrils diameter, according to Ding and Himmel’s model for maize 
biomass \cite{ding2006maize}. 

Further, height histograms of AFM images show a significant number of 
nanoelements with widths of less than the elementary fibril diameter 
[Fig.\,\ref{fig:afm-results}(b)]. This result suggests that in addition to 
disassembling the bundles, the oxidation may also weaken the noncovalent 
interchain and intersheet interactions within the elementary fibril, leading to 
the release of nanofibers with widths that correspond about to single (0.44\,nm) 
and double (0.88\,nm) cellulose polymer chains \cite{usov2015understanding}, as 
indicated by the arrows in Fig.\,\ref{fig:afm-results}(c). The width 
distribution may be an outcome of non-uniform distribution of charges along the 
fibers, as pointed out by Paajevan {\it et al.} \cite{paajanenCellulose2016}. 
\textcolor{black}{In addition to promoting the detachment of polymer chains, 
harsh oxidation reactions (SC-25 and SC-50) reduce the mass recovery ratio of 
TEMPO-oxidized pulps to ca 40\,\% compared to milder conditions (SC-5). This 
reduction is probably due to the removal of water-soluble molecules produced 
during oxidation, as discussed in our previous study\,\cite{pintoCarbPol2019}. 
Water-soluble products from TEMPO-oxidized cellulose pulps were also extracted 
by Hirota {\it et al.} using surface peeling with a solution of 20\,\% aqueous 
NaOH\,\cite{hirotaAngWantChem2010}.}

\subsection{Theoretical modeling}

\textcolor{black}{The present experimental results  suggest that the formation 
of 
cellulose polymer chains is a consequence of the weakening of (noncovalent) 
intersheet and interchain interactions within the elementary oxidized fibrils. 
Thus, in order to provide further support to these findings, we  have performed 
first-principles DFT simulations,  as described in Sec.\,II-B, addressing these 
noncovalent interactions. Firstly,  we} start our 
investigation with the pristine system, focusing on the role played by the 
hydrogen bonds and vdW forces in the strength of intersheet/interchain 
interactions. \textcolor{black}{In the sequence, we examine the changes of 
these 
IS and IC interactions upon the  presence of  charged  carboxylate groups 
embedded within  elementary  fibrils. We present an atomistic picture of 
how these charged carboxylate groups rule the \textcolor{black}{repulsive 
electrostatic forces} within the nanofibrils, which in its turn  weakens 
the noncovalent (IS and IC) interactions between the polymeric cellulose 
chains.}

\subsubsection{Pristine cellulose nanofibers}

In Fig.\,\ref{fig:models} we present the structural model of the monoclinic 
I$_\beta$ phase, where the periodic structure can be described by two misaligned 
molecular chains per unit-cell. The cellulosic sheets are composed of linear  
chains parallel to the {\bf a}  direction, where  the lateral interchain 
interaction ({\bf b} direction) are mostly ruled by  O--H$\cdots$O hydrogen  
bonds (HBs); while vdW interaction brings the main contribution to the 
structural stability  to  the sheet staking along the normal direction ({\bf 
c}). The strength of the intersheet (IS) and interchain (IC) interactions were 
quantified by the calculation of the IS and IC binding energies ($E^b_\text{IS}$ 
and $E^b_\text{IC}$, respectively),
\begin{eqnarray*}
 E^b_\text{IS} &=& E_\text{bulk}-E_\text{sheet} \\
 E^b_\text{IC} &=& E_\text{sheet}-E_\text{chain}.
\end{eqnarray*}
$E_\text{bulk}$, $E_\text{sheet}$, and $E_\text{chain}$ are the total energies 
of crystalline cellulose nanofibers, free-standing cellulosic sheet, and 
free-standing single molecular chain. The  binding energy of the cellulose 
nanofibers ($E^b_{\text{CNF}}$) can be written as:  
$E^b_{\text{CNF}}=E^b_\text{IC}+E^b_\text{IS}$. By using the vdW-DF 
approach\,\cite{thonhauserPRB2007,thonhauserPRL2015,berlandRepProgPhys2015, 
langrethJPhysC2009}, we found $E^b_{\text{CNF}}$\,=\,$-2.027$\,eV/unit-chain 
(unit-chain corresponds to two glucose rings)\,\footnote{Increasing the energy 
cutoff to 60\,Ry we obtained $E^b_\text{IS}=-1.321$\,eV/unit-chain and  
$E^b_\text{IC}=-0.665$\,eV/unit-chain, resulting in a binding energy of 
$-1.986$\,eV/unit-chain.}.  Our total energy results are summarized in Table\,I, 
where we have also considered the semi-empirical 
vdW-D2\,\cite{grimmeJCompChem2006} approach, to describe the long-range 
dispersive interactions. In Table\,II, we present  key informations  
regarding the equilibrium geometry of the I$_\beta$ crystalline cellulose, where 
we find a good agreement with the experimental measurements performed by 
Nishiyama {\it et al.}\,\cite{nishiyamaJACS2002}.

\begin{table}
\label{tab:energy}
\caption{Intersheet (IS), interchain (IC), and CNF binding energies for 
pristine  CNFs, using  vdW-DF, -D2 approaches implemented in the QE code,  and 
without vdW. The binding energies are in eV/unit-chain.}
\begin{ruledtabular}
\begin{tabular}{ccccc}
 vdW   & $E^b_\text{IS}$ & $E^b_\text{IC}$ &  $E^b_{\text{CNF}}$  &
$E^b_\text{IS}/E^b_\text{IC}$\\
\hline
DF          & $-1.351$  &  $-0.676$       &$-2.027$  &  2.0\\
D2       &   $-1.080$   &  $-0.833$       &$-1.913$  &  1.3  \\
no vdW   & $-0.127$  & $-0.569$      & $-0.697$     &   0.2 \\
\end{tabular}
\end{ruledtabular}
\end{table}

\begin{table}
\label{tab:geometry}
\caption{Equilibrium geometry of I$_\beta$ cellulose.}
\begin{ruledtabular}
\begin{tabular}{clll}
 vdW                  &     DF &     D2 & Exp.\,\cite{nishiyamaJACS2002} 
\\ 
\hline
  $a$ (\AA)           &  7.872 &  7.413 &  7.784      \\
  $b$ (\AA)           &  8.391 &  8.159 &  8.201     \\
  $c$ (\AA)           & 10.576 & 10.416 & 10.380      \\
  $\alpha$ ($^\circ$) & 90.0   &  89.8  & --          \\
  $\beta$  ($^\circ$) & 89.8   &  89.9  & --         \\
  $\gamma$ ($^\circ$) & 93.5   & 95.5   & 96.5        \\     
\end{tabular}
\end{ruledtabular}
\end{table}

Focusing on the IS and IC binding energies, there is a lack of consensus 
regarding the energetic balance between IS and IC iterations, where we may find 
different results obtained through different calculation approaches. For 
instance, Qian {\it et al.}\,\cite{qianMacromolec2005} found  IC interaction 
stronger than the IS interaction; while  Parthasarathi {\it et 
al.}\,\cite{parthasarathiJPhysChemA2011}  obtained nearly the same contribution 
for both interactions. Meanwhile, even upon the inclusion of vdW 
correction\,\cite{grimmeJCompChem2006},  Li {\it et 
al.}\,\cite{liJPhysChemC2011} obtained 0.8 and 1.1\,eV/unit-chain, for IS and IC 
interactions, respectively. In contrast, based on molecular dynamic (MD) 
simulations, Gross and Chu\,\cite{grossJPhysChemB2010,grossJPhysChemB2011} 
pointed out that the IS iteration is larger than IC interaction. Our total 
energy results support these latter findings. We found that  the IS interaction 
is stronger  than the IC one by almost twice, $E^b_\text{IS}/ 
E^b_\text{IC}\approx 2$ (vdW-DF results). By using the vdW-D2 approach, the 
strength of the IS (IC) interaction reduces (increases), however keeping 
$E^b_\text{IS}$ larger than $E^b_\text{IC}$ with  $E^b_\text{IS}/ 
E^b_\text{IC}\approx 1.3$. To check the accuracy of our results,  we have 
calculated the binding energies  using other nonlocal self-consistent vdW 
approaches. These results are summarized in Table\,III. 

As pointed out by Gross and Chu, vdW interaction rules the IS interaction, 
while it presents  a minor contribution to the IC one. In order to provide a 
quantitative picture of the role played by  the dispersive forces to the 
structural stability of the  I$_\beta$ CNFs, we calculate $E^b_\text{IS}$ and 
$E^b_\text{IC}$ by turning-off the vdW contribution, but keeping the 
equilibrium geometry obtained by the vdW-DF calculation. Here,  we found that 
the strength of the IS interaction reduces from $-1.351$ to 
$-0.127$\,eV/unit-chain, while for the IC interactions change by less than 
0.1\,eV,  $E^b_\text{IC}=-0.676\rightarrow-0.569$\,eV/unit-chain. Thus, 
revealing that   vdW forces present (i) a dominant role on the energetic 
stability between the cellulosic sheets (about 90\,\% of the IS interactions), 
and (ii) a minor contribution to the IC interactions ($\sim$16\,\%). These 
findings [(i) and (ii)] will be helpful to provide an energetic picture for the 
disintegration of oxidized CNFs as discussed below. 

\hiroki{We are aware that  the presence of solvents may  contribute to  the 
energetic stability involving cellulose interfaces. For instance, the role 
played by  water molecules  on the energetic stability of cellulose/lipid 
interfaces\,\cite{gurtovenko2018phospholipid, gurtovenko2019controlled}, and 
the reorganization processes of cellulose
sheets\,\cite{miyamoto2009structural,miyamoto2013structural}.
In the present study  we have not included the contribution of 
solvent in the calculations of the binding energies. Here,  $E^b_\text{IC}$ and 
$E^b_\text{IS}$ were obtained comparing the total energies, at $T$=0, of the  
initial (chain/sheet) and final (sheet/bulk) ground state configurations.  We 
are assuming the  approximation that (i) the solvent  contribution to the   
total energies of the  initial and final systems  are nearly the same,  as well 
as (ii)  the temperature dependence carried out by the entropic term in the free 
energy. Therefore,  within such an approximation, these contributions [(i) 
and (ii) separately] are canceled out  when we compare the initial 
and final free energies\,\cite{qian1988first,deshpande2008binding}.}

\subsubsection{Oxidized cellulose nanofibers}

The energy cost to disintegrate the cellulose fibers reduces upon the formation 
of charged carboxylate groups (COO$^-$)\,\cite{tejadoCellulose2012}. Indeed, 
previous experimental works have  shown the disintegration of cellulose  into 
nanofibers with diameters of 3-5\,nm mediated by (TEMPO) oxidation  
processes\,\cite{saitoBiomacro2004,saitoBiomacro2006}. In parallel,   MD 
simulations have been done addressing the effect of the presence of carboxylate 
groups, bonded to the CNF surfaces, on the inter-fibril 
interaction\,\cite{paajanenCellulose2016}. Here, we have used a recently 
proposed  low energy cost pathway\,\cite{pintoCarbPol2019}  to produce cellulose 
nanofibers with widths of single and double cellulose polymer chains, as 
depicted in Fig.\,\ref{fig:afm-results}.

\begin{table}
\label{tab:energy2}
\caption{Intersheet (IS), and interchain (IC), and CNF binding energies using 
the vdW-DF, -DF2, and -optB86b approaches 
implemented in the VASP code. The binding energies are in eV/unit-chain.}
\begin{ruledtabular}
\begin{tabular}{ccccc}
   vdW   & $E^b_\text{IS}$ & $E^b_\text{IC}$ &  $E^b_\textrm{CNF}$ & 
$E^b_\text{IS}/E^b_\text{IC}$\\
\hline
DF         & $-1.288$   & $-0.663$    & $-1.951$ & 1.9  \\  
DF2        & $-1.232$   & $-0.716$    & $-1.948$ & 1.7 \\
optB86b    & $-1.441$   & $-0.818$    & $-2.259$ & 1.8 \\

\end{tabular}
\end{ruledtabular}
\end{table}

  \begin{figure}
    \includegraphics[width=6.5cm]{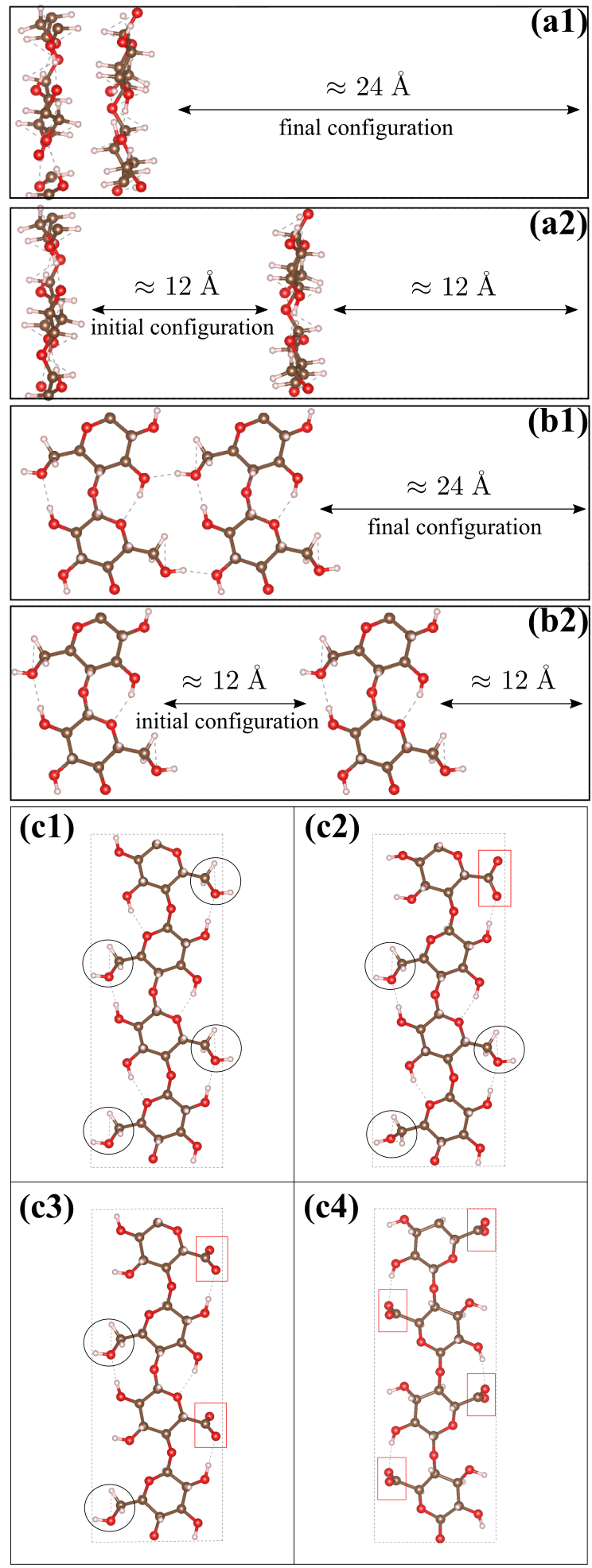}
    \caption{\label{fig:calc-scheme} Schematic representation of the IS (a) and 
IC (b) binding energy calculations.  (a1)-(b1)/(a2)-(b2) Final/Initial 
configuration  of two free standing cellulosic  sheets\,(a) and chains\,(b).
(c) Structural models of pristine 
(non-oxidized) NC chain (c1), and oxidized chains with linear 
concentration of carboxylate groups of [COO$^-$] = 25\% (c2), 50\% (c3), and 
100\% (c4). The carboxylate groups are within rectangles.} 
  \end{figure}  
  
  \begin{figure}
    \includegraphics[width=\columnwidth]{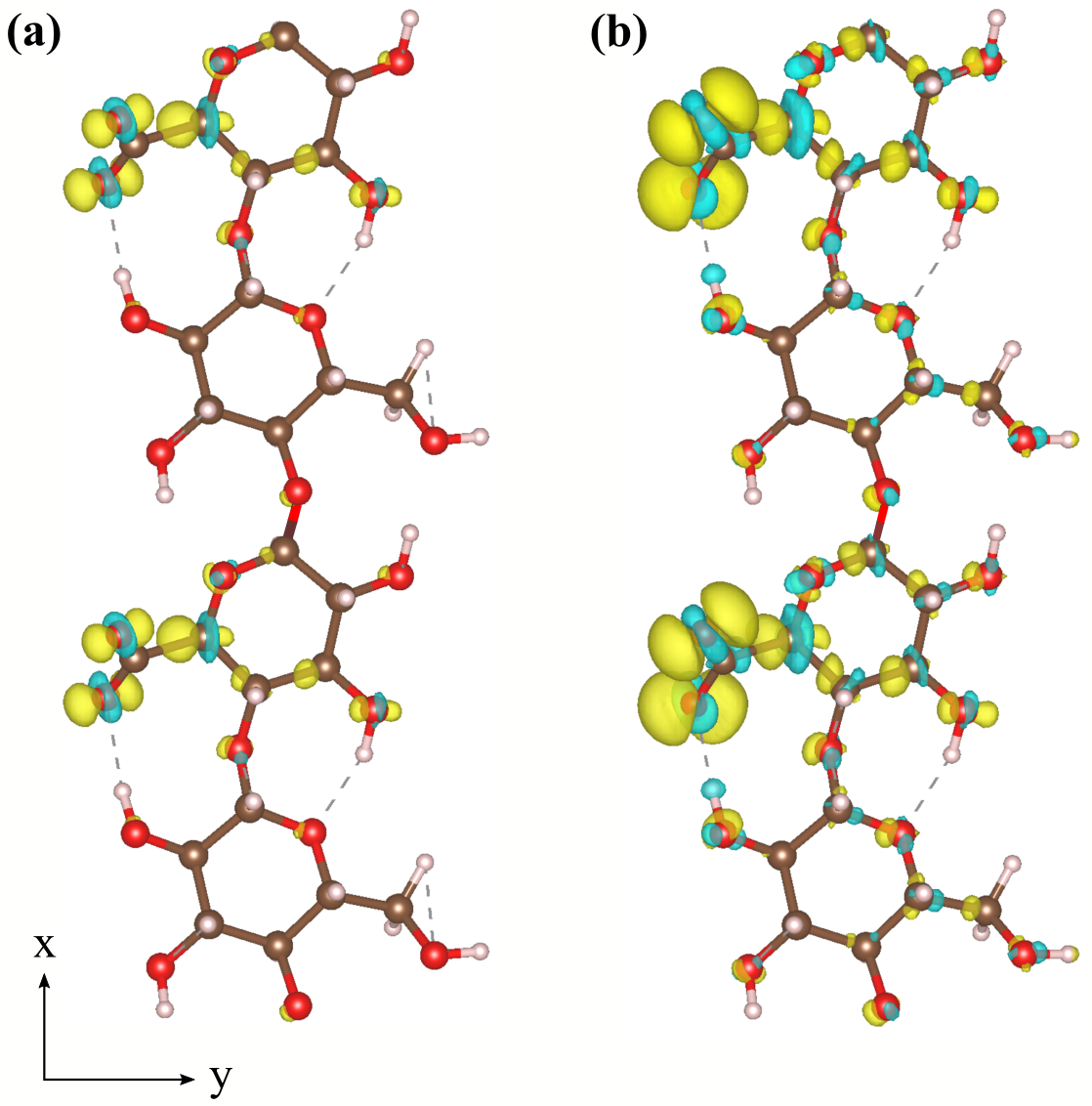}
    \caption{\label{delta-rho} Charge density distribution along 
oxidized cellulose  chain, with [COO$^-$] of 50\%, upon the net charging 
increase $q$=0.25\,$\rightarrow$\,0.50\,$e$ (a) and 
$q$=0.25\,$\rightarrow$\,1.0\,$e$ (b). Isosurfaces of 0.003 {\itshape 
e}/\AA$^{3}$ in (a) and 0.002 {\itshape e}/\AA$^{3}$ in (b).}
  \end{figure} 
  
Given such a scenario, and the present experimental findings, it is expected 
that the formation of carboxylate groups should take place not only on the  
surface of the elementary fibrils, but also at  the inner sites of the CNFs 
[indicated by dashed lines in Fig.\,\ref{fig:afm-results}(d)],  weakening the 
intersheet and interchain  interactions.

The calculation procedure of the IS and IC binding energy, as a function of the 
charging state  of the carboxylate group,  is  schematically shown in 
Figs.\,\ref{fig:calc-scheme}(a) and (b). Here, the binding energy ($E^b$) 
corresponding to the IS [IC] interaction was obtained by  comparing the total 
energies of two free-standing cellulosic sheets [chains] interacting to each 
other (final configuration), as shown in  Fig.\,\ref{fig:calc-scheme}(a1) 
[\ref{fig:calc-scheme}(b1)], and the ones far from each other (initial 
configuration),  Fig.\,\ref{fig:calc-scheme}(a2)  [\ref{fig:calc-scheme}(b2)], 
for a given charging state ($q$)\,\footnote{We check the convergence of our 
results with respect to the size of the supercell, since in the charged systems 
a jellium background of opposite charge has been inserted in order to work with 
neutral supercells. Here, we found that by increasing the vacuum region from 24 
to 40\,\AA\, our results of $E^b(q)$ change by $\sim$4\%.},  
$$ 
E^b_{\rm IS/IC}(q)=E_{\textrm{final}}(q) -  E_{\textrm{initial}}(q).
$$ 
For each configuration (initial/final), the fully relaxed atomic positions,
and  the total energies were obtained including the vdW interactions within the 
vdW-DF approach\cite{thonhauserPRL2015,thonhauserPRB2007,berlandRepProgPhys2015, 
langrethJPhysC2009}. Within our supercell approach, we have considered the 
presence  of carboxylate groups with different concentrations ([COO$^-$]),  as 
shown in Figs.\,\ref{fig:calc-scheme}(c1)-(c4). The localization of the net 
charge ($q$) was determined by using the L$\rm\ddot{o}$wdin orbital 
population\,\cite{lowdin},  where we found that most of charging lies on the 
carboxylate groups. In Figs.\,\ref{delta-rho}(a) and (b), we present the 
increase of the net charge density along a cellulose chain with [COO$^-$] of 
50\% when the charging state increases from $q$=0.25\,$\rightarrow$\,0.50\,$e$ 
and $q$=0.50\,$\rightarrow$\,1.00\,$e$. To check the adequacy of our calculation 
approach, instead of charging our supercell,   we have considered the presence 
of sodium ions nearby the carboxylate groups\,\cite{paajanenCellulose2016}. In 
this case, we found a net charging of about 0.5\,$e$ per [COO$^-$] unit. 
 \textcolor{black}{Here, we have not considered the disordered  structures 
present at the edge sites of the CNFs. Eventually, the formation of  
carboxylate groups at the edge sites may trigger the disassembling 
process. However, the role played by the oxidized sites, weakening 
the IC and IS interaction, should be the same as those we have predicted for 
the inner sites of the CNFs.}
  
  \begin{figure}
    \includegraphics[width=\columnwidth]{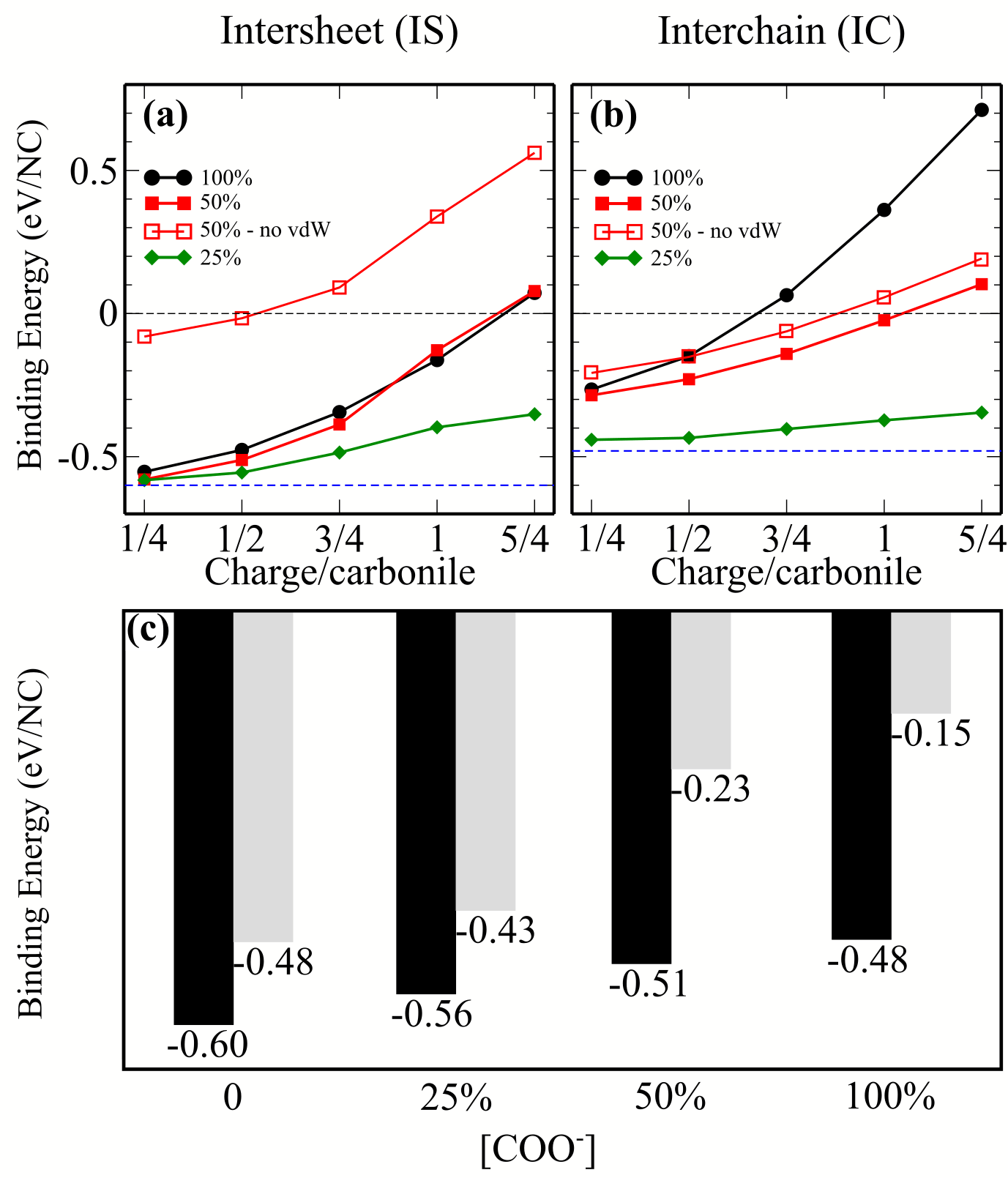}
    \caption{\label{binding-ox} Intersheet (a) and interchain (b) binding energy
    as a function of the charging state ($q$) with [COO$^-$]=100, 50, and 
25\%. Dashed lines indicate the calculated binding energies of a 
(non-oxidized) pristine system. (c) IS (black) and IC (shaded) binding 
energies for $q=0.5 e$. [COO$^-$]\,=\,0 indicates non-oxidized pristine system.}
  \end{figure}  
  
Our results of binding energies  for the IS and IC interactions 
[Figs.\,\ref{binding-ox}(a) and (b)]   reveal that the weakening of the  IS and 
IC interactions, compared with the ones of pristine systems (dashed lines in 
Fig.\,\ref{binding-ox}), is proportional to the charging and the concentration 
of carboxylate groups, \textcolor{black}{which is  in line with 
our experimental findings. As shown in  Section III-A, the formation of 
cellulose polymer chains takes place  in SC-25 and -50,  but not in SC-5 due to 
its low concentration of charged carboxylated sites.
Moreover, the  net (negative) charge localization in the 
carboxylate group, as shown in Fig.\,\ref{delta-rho}, indeed, support  the  
repulsive electrostatic role on the cellulose disassembling process of 
CNF by weakening  the IS and IC interactions\,\cite{tejadoCellulose2012}.}

Further binding energy comparison [Fig.\,\ref{binding-ox}(c)] shows that, in 
general,  the IC interactions are more sensitive to the oxidation than the IS 
interactions. For instance, for [COO$^-$]=50\% and charging of $q=0.5 e$ the IC 
binding energy reduces by 0.25\,eV/unit-chain, $E^b_{\rm IC}= -0.48\rightarrow 
-0.23$\,eV/unit-chain, whereas for the IS interaction we found $E^b_{\rm IS}$ 
reduces by 0.09\,eV/unit-chain, $-0.60\rightarrow -0.51$\,eV/unit-chain. These 
results not only support the present experimental findings (summarized in 
Fig.\,\ref{fig:afm-results}) but also  allow us to infer that  the CNF 
disassembly should take place (preferentially) through a  disruption of the 
interchain interactions [indicated by red dashed lines in 
Fig.\,\ref{fig:afm-results}(d)].

As we have discussed above, in the pristine systems, the IS binding energy is 
mostly dictated by  vdW forces, while the hydrogen-like C--H$\cdots$O bonds 
bring a minor contribution. In contrast, the IC binding strength is mostly ruled 
by  O--H$\cdots$O hydrogen bonds, followed by  a minor contribution from vdW 
forces.  Since, in the oxidized systems,   $E^b_{\rm IC}$ presents greater 
reduction compared with $E^b_{\rm IS}$ [Fig.\,\ref{binding-ox}(c)], here we can 
infer that the weakening  O--H$\cdots$O and C--H$\cdots$O bonds  play the main 
role in the disassembly process of  CNFs. In order to provide a qualitative 
support to such a statement, we calculate $E^b_{\rm IS/IC}(q)$,   for 
[COO$^-$]=50\%, with no vdW contribution to the total energies. Our results for 
the IS and IC binding energies [indicated by empty squares in 
Figs.\,\ref{binding-ox}(a) and (b)]  reveal a significant reduction of the 
former. For instance, for $q=0.5\,e$, $E^b_{\rm IS}(q)$ reduces from $-0.511$ to 
$-0.016$\,eV/unit-chain, whereas  the strength of the IC interaction reduces by 
about 0.08\,eV/unit-chain, $E^b_{\rm IC}(q)=-0.230\rightarrow 
-0.151$\,eV/unit-chain. In other words, the higher reduction of $E^b_{\rm IS}$ 
compared with that of $E^b_{\rm IC}$ is a consequence of the predominance of van 
der Waals forces in the intersheet interactions. 

\hiroki{We have used the current state-of-the-art DFT-vdW approach to describe 
the non-covalent interactions\,\cite{stohr2019theory,kim2020establishing}, 
however, there are other ``ingredients'' that we have not considered  in the 
present study of oxidized CNFs. (i) Water molecules; the preferential 
incorporation of  water molecules at the hydrophilic regions, of the CNFs, may 
favor the formation of cellulose chains instead of 
sheets\,\cite{zhang2020polarities}.  (ii) Hydrated (positively charged) 
counter-ions, like 
[Na(H$_2$O)$_n$]$^+$\,\cite{deshpande2008binding,heyda2012ion, 
paajanenCellulose2016,peng2018effect,qi2020cooperative}, resulting in a 
screening effect of  the repulsive electrostatic interaction between the 
negatively charged carboxylate groups. Further investigations are necessary 
including (i) and (ii), combined with an accurate description of the noncovalent 
vdW interactions, in order to provide a more complete picture of the formation 
of cellulose polymer chains. However, despite the lack of these ``ingredients'', 
we believe that our simulations results are valuable for providing not only 
theoretical support to the experimental findings, but also  an atomic scale 
understanding of the disassembling processes in oxidized CNFs.}

\section{Summary and Conclusions}

We have performed an experimental and theoretical combined investigation of 
cellulose disassembly  mediated  oxidation processes. Based on a low energy cost 
pathway, single and double cellulose polymers chains have been synthesized from 
oxidized cellulose nano fibers (CNFs). Such a disruption of CNFs was attributed 
to the formation of carboxylate groups embedded  within oxidized fibrils, 
weakening the noncovalent interchain (IC) and intersheet (IS) interactions.  
\textcolor{black}{We have shown  that the disassembling process 
depends on the concentration of the carboxylate groups.} 
First-principles DFT calculations revealed that, indeed,  the IC and IS binding 
energies [$E^b_{\rm IC/IS}(q)$] reduce in the oxidized CNFs.  We find that such 
a  reduction of $E^b_{\rm IC/IS}(q)$ is proportional to the charging state 
($q$), \textcolor{black}{and the concentration of the oxidized carboxylate 
groups} indicating an electrostatic repulsion between the (charged) cellulose 
fibrils. Finally, comparing the IS and IC binding energies, we found that the 
disassembly processes of the oxidized CNFs should take place mostly through a 
disruption of the  interchain interactions, giving rise to (predominantly) 
fibers and cellulose chains, instead of sheets, as a final structure, 
\textcolor{black}{and thus in occordance/supporting with the present 
experimental 
AFM observations.}

%
%

\begin{acknowledgments}

The authors acknowledge   financial   support   from   the Brazilian  agencies  
CNPq, FAPEMIG, and FAPESP (grant 16/04514-7 and 17/02317-2), and the  LNCC 
(SCAFMat2), CENAPAD-SP for 
computer time.

\end{acknowledgments}

\bibliography{RHMiwa-cellu}
\end{document}